\documentstyle[12pt,fleqn,epsf]{article}
\def\nsection#1{\setcounter{equation}{0}\section{#1}}

\newcommand{\Integer}{\:\mbox{\sf Z} \hspace{-0.82em} \mbox{\sf Z}\,}
\newenvironment{eqar}{
        \setlength{\mathindent}{0.7 cm}\begin{eqnarray}}{\end{eqnarray}}
\newenvironment{eqr}{
        \setlength{\mathindent}{0 cm}\begin{eqnarray}}{\end{eqnarray}}
\def\Mult#1#2#3{\left[#1 \atop #2 \right]_{#3}}
\def\Mults#1#2#3{\left[{\textstyle {#1 \atop #2} } \right]_{#3}}
\def\la{\lambda}

\def\e{\mbox{e}}
\def\vec#1{\mbox{\boldmath $#1$}}
\def\svec#1{\mbox{\scriptsize \boldmath $#1$}}

\begin{document}

\title{Exceptional structure of the dilute A$_3$ model:
\\ E$_8$ and E$_7$ Rogers--Ramanujan identities}

\author{S.~Ole Warnaar\thanks{
e-mail: {\tt warnaar@mundoe.maths.mu.oz.au}}
\ and Paul A.~Pearce\thanks{
e-mail: {\tt  pap@mundoe.maths.mu.oz.au}} \\
Mathematics Department\\
University of Melbourne\\
Parkville, Victoria 3052\\
Australia}


\maketitle

\begin{abstract}
The dilute A$_3$ lattice model in regime 2 is in the universality
class of the Ising model in a magnetic field. Here we establish
directly the existence of an E$_8$ structure in the dilute A$_3$ model
in this regime by expressing the 1-dimensional configuration sums in
terms of fermionic sums which explicitly involve the  E$_8$ root system.
In the thermodynamic limit, these polynomial identities yield a proof
of the E$_8$ Rogers--Ramanujan identity recently conjectured by
Kedem {\em et al}.
The polynomial identities also apply to regime 3, which is obtained by
transforming the modular parameter by $q\to 1/q$. In this case we find
an A$_1\times\mbox{E}_7$ structure and prove a Rogers--Ramanujan
identity of A$_1\times\mbox{E}_7$ type. Finally, in the critical
$q\to 1$ limit, we give some intriguing  expressions for the number
of $L$-step paths on the A$_3$ Dynkin diagram with tadpoles in
terms of the E$_8$ Cartan matrix. All our findings confirm the
E$_8$ and E$_7$ structure of the dilute A$_3$ model found recently
by means of the thermodynamic Bethe Ansatz.
\end{abstract}

\nsection{Introduction}
Recently, a Bethe Ansatz study \cite{BazNienWar} of the
dilute A$_3$ lattice model \cite{WarNienSea,WarPeaSeaNien}
has revealed a hidden E$_8$ structure.
This establishes the expected relation between the dilute A$_3$ model
and Zamolodchikov's E$_8$ $S$-matrix of the critical Ising
model in a field \cite{Zam}. One of the drawbacks, however, of this
Bethe Ansatz approach is that it relies heavily on the acceptance of a
conjectured string structure of the Bethe Ansatz equations.
In this letter we demonstrate the E$_8$ structure of the dilute A$_3$ model
directly, without the use of a string hypothesis.

In recent papers Melzer~\cite{Melzer} and  Berkovich \cite{Berk} have
shown that the 1-dimensional configuration sums of the ABF model
admit a so-called fermionic representation in addition to the
well-known bosonic forms of Andrews {\em et al.}~\cite{ABF}.
Their motivation was in fact to prove Rogers--Ramanujan (RR)
type identities for the $\chi_{r,s}^{(h)}$ Virasoro characters
associated with the minimal unitary models of central
charge $c=1-6/h(h-1)$ as conjectured by the Stony Brook group~\cite{Kedem1}.
In this letter we adopt a similar approach. Specifically,
we rewrite the known bosonic expressions for the
1-dimensional configuration sums of the dilute A$_3$ model
in regime 2$^{+}$ \cite{WarPeaSeaNien} in terms of fermionic sums.
These fermionic sums explicitly involve the E$_8$ root system.
In particular, in the thermodynamic limit,
our ``fermionic sum = bosonic sum'' expressions yield
precisely the E$_8$ Rogers--Ramanujan identity for the
$\chi_{1,1}^{(4)}$ Virasoro character as given by
Kedem {\em et al.}~\cite{Kedem2}.

Thermodynamic Bethe Ansatz computations were also
carried out~\cite{BazNienWar} in regime
3$^{+}$ of the dilute A$_3$ model. In this case the model is
known~\cite{WarPeaSeaNien} to decouple in the scaling limit into an Ising
model and a $\phi_{2,1}^{(5)}$ perturbed minimal model. Accordingly,
the Bethe Ansatz computations~\cite{BazNienWar,BazResh} on the
dilute A$_3$ model in regime 3$^{+}$ gives an A$_1\times\mbox{E}_7$
structure leading to the correct central charge $c=1/2 + 7/10 = 6/5$.
These findings for regime 3$^+$ are supported in this letter.
By considering the thermodynamic limit of our
``fermionic sum = bosonic sum'' expressions, after
carrying out the transformation $q \to 1/q$, which maps
regime 2$^{+}$ onto regime 3$^{+}$ \cite{WarPeaSeaNien},
we find an A$_1\times\mbox{E}_7$ Rogers--Ramanujan identity
similar to the E$_7$ identity for the $\chi_{1,1}^{(5)}$
character conjectured by Kedem {\em et al.}~\cite{Kedem2}.

To conclude this letter, we point out some intriguing expressions
for the number of walks on the adjacency graph of the dilute A$_3$ model
in terms of the E$_8$ Cartan matrix.

\nsection{Polynomial E$_8$ Rogers--Ramanujan identity}
Before we present the main results of this letter we
need to introduce some notation.

We define the Gaussian multinomials or $q$-multinomials
by \cite{Andrews}
\begin{equation}
\Mult{N}{m_1,m_2,\ldots,m_n}{q} = \frac{(q)_N}{(q)_{m_1} (q)_{m_2}
\ldots (q)_{m_n} (q)_{N-m_1-m_2-\ldots -m_n}} \; ,
\end{equation}
where $(q)_m = \prod_{k=1}^m (1-q^k)$ for $m>0$ and $(q)_0=1$.
Also, if  ${\cal I}_{\mbox{\scriptsize E}_8}$ denotes the
incidence matrix of E$_8$ with the nodes labelled as in figure~1a,
we define the following thermodynamic Bethe Ansatz (TBA) type systems:
\begin{equation}
\vec n + \vec m =\frac{1}{2} \left ( {\cal I}_{\mbox{\scriptsize E}_8}
\vec m + (L-1) \vec e_1 +\vec e_i \right) \qquad i=1,2,\ldots, 8.
\label{TBA}
\end{equation}
Here $\vec n$ and $\vec m$ are 8-dimensional column vectors with
{\em integer}  entries $n_i$, $m_i$, respectively,
and $\vec e_i$ is a unit vector with components
$(\vec e_i)_j=\delta_{i,j}$.
The parameter $L$ will be referred to as the system size.
A pair of vectors $\vec{n}$ and $\vec{m}$ solving the $i$-th TBA type
equation with system size $L$ will be denoted by
$(\vec{n},\vec{m})_{L,i}$.
Following ref.~\cite{Berk}, we now define the fermionic
functions $F_i(L)$ by
\begin{equation}
F_i(L) = \sum_{(\svec{n},\svec{m})_{L,i}} q^{\svec{n}^T
C_{\mbox{\tiny E}_8}^{-1} \, \svec{n} }
\prod_{j=1}^8
\Mult{n_j + m_j}{n_j}{q},
\end{equation}
with $C_{\mbox{\scriptsize E}_8}$ the Cartan matrix of E$_8$ which is
related to the incidence matrix by
$(C_{\mbox{\scriptsize E}_8})_{i,j}= 2 \delta_{i,j} -
({\cal I}_{\mbox{\scriptsize E}_8})_{i,j}$.
Finally, we define the bosonic functions
$B_{r,s}(L,a,b)$ by \cite{WarPeaSeaNien}
\begin{eqnarray}
B_{r,s}(L,a,b) &=& \sum_{j,k=-\infty}^{\infty}
\left\{q^{12j^2+(4r-3s)j+k(k+8j+b-a)}
\Mult{L}{k,k+8j+b-a}{q} \right.
\nonumber \\
& & \qquad  \left. -q^{12j^2+(4r+3s)j+rs+k(k+8j+b+a)}
\Mult{L}{k,k+8j+b+a}{q} \right\}.
\end{eqnarray}
With these definitions our main assertion can be written as
\begin{equation} F_1(L) = B_{1,1}(L,1,1).
\end{equation}
Explicitly, this polynomial identity takes the form
\begin{eqnarray}
\lefteqn{
\sum_{(\svec{n},\svec{m})_{L,1}}
q^{\svec{n}^T C_{\mbox{\tiny E}_8}^{-1} \, \svec{n} }
\prod_{i=1}^8 \Mult{n_i + m_i}{n_i}{q} } \nonumber \\
& &= \sum_{j,k=-\infty}^{\infty}
\left\{q^{12j^2+j+k(k+8j)}
\Mult{L}{k,k+8j}{q}  -q^{12j^2+7j+1+k(k+8j+2)}
\Mult{L}{k,k+8j+2}{q} \right\},
\label{E8id}
\end{eqnarray}
which can be viewed as a finitization of the
E$_8$ Rogers--Ramanujan
identity of Kedem {\em et al.}~\cite{Kedem2}.
Indeed, taking the limit $L\to\infty$,
using the result \cite{WarPeaSeaNien}
\begin{equation}
\lim_{L\to\infty} \sum_{k=-\infty}^{\infty} q^{k(k+a)}
\Mult{L}{k,k+a}{q} = \frac{1}{(q)_{\infty}} \: ,
\end{equation} together with the simple formula
$\lim_{N\to\infty}\Mult{N}{m}{q}  = 1/(q)_m$, gives
\begin{equation}
\sum_{n_1,\ldots,n_8=0}^{\infty}
\frac{q^{\svec{n}^T C_{\mbox{\tiny E}_8}^{-1}\, \svec{n} } }
{(q)_{n_1} \ldots (q)_{n_8} }  =
\frac{1}{(q)_{\infty}}
\sum_{j=-\infty}^{\infty} \left\{
 q^{12j^2+j}-q^{12j^2+7j+1} \right\} .
\end{equation}
The RHS of this E$_8$ Rogers--Ramanujan identity is the
usual (bosonic) Rocha-Caridi form for the $\chi_{1,1}^{(4)}$
Virasoro character~\cite{Rocha}.
The LHS is the fermionic counterpart conjectured by
Kedem {\em et al.}~\cite{Kedem2}.

Before we proceed to sketch a proof of identity (\ref{E8id}),
let us first explain how the above results relate to the E$_8$
structure of the dilute A$_3$ model.
To this end we note that the bosonic side of equation (\ref{E8id})
is precisely the expression for the 1-dimensional configuration
sum $Y^{1\,1\,1}_L(q)$  of the dilute A$_3$ model in regime 2$^{+}$
as computed in \cite{WarPeaSeaNien}.
More generally, the configuration sums $Y^{a\,b\,c}_L(q)$ with
$a,b,c\in\{1,2,3\}$ and $|b-c|\le 1$ are defined via
\begin{equation}
Y_L^{\sigma_1\,\sigma_{L+1}\,\sigma_{L+2}}(q)
=\sum_{\sigma_2,\ldots,\sigma_L} q^{\sum_{j=1}^L j
H(\sigma_j,\sigma_{j+1},\sigma_{j+2})}.
\end{equation}
The function $H$ herein follows directly from the Boltzmann weights
of the dilute A$_3$ model by computing the ordered infinite field
limit $(p\to 1$, $u/\epsilon$ fixed)
\begin{equation}
W \left(
\begin{array}{cc} d & c \\ a & b \end{array}
\right) \to \frac{g_a g_c}{g_b g_d} \;
\e^{-2 \pi u H(d,a,b)/\epsilon} \:
\delta_{a,c} \qquad \mbox{with}
\qquad g_a = \e^{-2\la u a^2/\epsilon}.
\end{equation}
Here $u$ is the spectral parameter, $3 \la=15\pi/16$ is the crossing
parameter and $p=\exp(-\epsilon)$ is the nome of the elliptic function
parametrization of the face weights~\cite{WarPeaSeaNien}.
A complete listing of the values of $H(a,b,c)$ is given in (A.5)
of ref.~\cite{WarPeaSeaNien}.
The occurrence of the particular configuration sum $Y^{1\,1\,1}_L(q)$ in
(\ref{E8id}) can be understood from the fact that the configuration  with
all spins on the lattice taking the value 1 corresponds to the ground
state of the model.

\begin{figure}[hbt]
\centerline{\epsffile{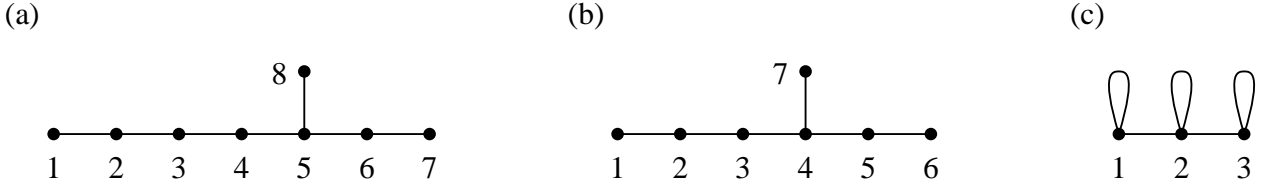}}
\caption{(a) The Dynkin diagram of E$_8$. (b) The Dynkin diagram of E$_7$.
(c) The incidence or adjacency graph of the dilute A$_3$ model.}
\end{figure}

\subsection{Sketch of a proof of (2.7)}
The proof of identity (\ref{E8id}) is long and tedious so we will
present it in full elsewhere.
Here we only indicate the main ingredients of the proof and omit
the detailed calculations.

Let us recall that the configuration sums
$Y^{a\,b\,c}_L(q)$ satisfy the recurrences \cite{WarPeaSeaNien}
\begin{equation}
Y_L^{a\,b\,c}(q)=q^{L\, H(b-1,b,c)} \, Y_{L-1}^{a\,b-1\,b}(q)
+q^{L \, H(b,b,c)} \, Y_{L-1}^{a\,b\,b}(q)
+q^{L\,  H(b+1,b,c)} \, Y_{L-1}^{a\,b+1\,b}(q)
\label{recur}
\end{equation}
subject to the initial condition
\begin{equation}
Y_1^{a\,b\,c}(q)=q^{H(a,b,c)}.
\label{init}
\end{equation}
Moreover, the recurrence relations together with the initial
conditions uniquely determine the configuration
sums $Y^{a\,b\,c}_L(q)$. In view of (\ref{E8id}) we
will only consider the case $a=1$.
Apart from $Y^{1\,1\,1}_L(q)$ we now also express the other
$Y^{1\,b\,c}_L(q)$ firstly in terms of fermionic sums and secondly
in terms of bosonic sums.
If we can then show that both the fermionic and bosonic expressions
satisfy the recurrences (\ref{recur}) together with (\ref{init}),
the fermionic sums must equal the bosonic ones due to the
uniqueness of solution to (\ref{recur}) and (\ref{init}).

First, we list the bosonic expressions for $Y^{1\,b\,c}_L(q)$:
\begin{eqnarray}
Y^{1\,1\,1}_L(q) &=& B_{1,1}(L,1,1)  \nonumber \\
Y^{1\,1\,2}_L(q) &=& q^L B_{1,1}(L,1,1)  \nonumber \\
Y^{1\,2\,1}_L(q) &=& B_{3,1}(L,1,2) \nonumber
\\ Y^{1\,2\,2}_L(q) &=& q^{L-1} B_{1,1}(L,1,2)  + q^L(1-q^L)
B_{3,1}(L-1,1,3) \\
Y^{1\,2\,3}_L(q) &=& q^{2L-1} B_{1,1}(L,1,2) \nonumber \\
Y^{1\,3\,2}_L(q) &=& q^{-L-1} B_{3,1}(L,1,3) \nonumber \\
Y^{1\,3\,3}_L(q) &=& q^{L-1} B_{3,1}(L,1,3)
+ q^{L-3} (1-q^L) B_{1,1}(L-1,1,2). \nonumber
\end{eqnarray}
The proof that this solves (\ref{recur}) and (\ref{init})
has been given in \cite{WarPeaSeaNien}.

Second, we list fermionic expressions for $Y^{1\,b\,c}_L(q)$:
\begin{eqar}
Y^{1\,1\,1}_L(q) &=& F_1(L) \nonumber \\
Y^{1\,1\,2}_L(q) &=& q^L F_1(L) \nonumber \\
Y^{1\,2\,1}_L(q) &=& \left( F_7(L)-(1-q^L)F_1(L)
-q^L(1-q^L)F_1(L-1) \right. \nonumber \\
& & \left.
+q^{2L-1}(1-q^{L-1})F_1(L-2) \right)/q^{L+1} \nonumber \\
Y^{1\,2\,2}_L(q) &=& \left( F_7(L)-(1-q^L)F_1(L)
+q^{2L-1}(1-q^{L-1})F_1(L-2) \right)/q  \nonumber \\
Y^{1\,2\,3}_L(q) &=& q^{2L-1} \left( F_7(L)
+ q^{L-1}(1-q^{L-1}) F_1(L-2)\right) \label{ferm1pf} \\
Y^{1\,3\,2}_L(q) &=& \left( F_7(L)-F_1(L)+q^L F_7(L-1)
+q^{L-1}(1-q^{L-1})F_7(L-2) \right. \nonumber \\
& & \left.  +q^{2L-2}(1-q^{L-2})F_1(L-3)
+q^{2L-4}(1-q^{L-1})(1-q^{L-3})F_1(L-4) \right)/q^{L+3} \nonumber \\
Y^{1\,3\,3}_L(q) &=& q^{L-3}
\left(F_7(L)-F_1(L)+F_7(L-1)
+q^{L-1}(1-q^{L-1})F_7(L-2) \right.  \nonumber \\
& & \left.  +q^{L-2}(1-q^{L-2})F_1(L-3)
+q^{2L-4}(1-q^{L-1})(1-q^{L-3})F_1(L-4) \right).  \nonumber
\end{eqar}
These expressions are admittedly quite complicated and might very
well be simplified.  For example, we have chosen to express all
configuration sums in terms of $F_1$ and $F_7$ only.
Using easily verifiable recurrences of the type
$F_1(L)-q^{2L-2} F_1(L-2) = F_2(L-1)$,
$F_2(L)+q^{L+1}(1-q^{L-2}) F_2(L-2) = F_3(L-1) + q^{L+1} F_1(L-1)$, etc.,
one could conceivably find simpler forms for the above.

To prove the correctness of the fermionic solution to the recurrence
relations we substitute (\ref{ferm1pf}) into (\ref{recur}).
This gives seven identities which
can be combined to yield the following two equations
\begin{eqr}
\lefteqn{F_1(L) -F_7(L-1) -q^{L-1}F_1(L-1) +q^{L-1}(1-q^{L-1})F_1(L-2)
-q^{2L-3}(1-q^{L-2})F_1(L-3) = 0} \nonumber \\
\lefteqn{F_7(L) -q^2 F_1(L-1) -(1+q^{L-1})F_7(L-1) +q^2
(1-q^{2L-4})F_7(L-2) } \\
& & \quad +q^{L-1}F_1(L-2)   -q^L (1-q^{L-2})F_1(L-3) +q^L
(1-q^{L-2})(1-q^{L-3})F_7(L-3) = 0. \nonumber
\end{eqr}
The actual proof of these final two equations will be omitted
here, but we remark that they follow from elementary but
tedious computations very similar to those carried out in \cite{Berk}.
The proof that (\ref{ferm1pf}) satisfies the
initial conditions is a matter of straightforward case checking.

\nsection{A$_1\times\mbox{E}_7$ Rogers--Ramanujan identity}
In this section we consider the $L\to\infty$ limit of
identity (\ref{E8id}) after first  replacing $q$ by $1/q$.
The effect of this transformation on $q$ is to map
from regime 2$^+$ to regime 3$^+$ of the dilute A$_3$ model.
The critical behaviour of the model in this latter regime is
described by \cite{WarPeaSeaNien,BazNienWar} a $c=6/5$
conformal field theory given as a direct product of an
Ising model ($c=1/2$) and an E$_7$ theory with $c=2
\; \mbox{rank} \; {\cal G}/(g+2)= (2)(7)/(18+2)=7/10$.
Hence it is to be expected that the above steps will
result in an A$_1\times\mbox{E}_7$ Rogers--Ramanujan identity.

To establish this we use two simple inversion formulas:
\begin{eqnarray}
\Mult{N}{m}{1/q} &=& q^{m(m-N)} \Mult{N}{m}{q} \label{inv2} \\
\Mult{N}{m_1,m_2}{1/q} &=& q^{m_1^2+m_2^2+m_1 m_2-(m_1+m_2)N}
\Mult{N}{m_1,m_2}{q} . \label{inv3}
\end{eqnarray}
Applying (\ref{inv3}) to transform the bosonic RHS of (\ref{E8id})
we obtain, after performing a shift on the summation variable $k$,
\begin{eqnarray}
\lefteqn{q^{(\mu-L^2)/2} \sum_{j,k=-\infty}^{\infty} q^{2k(k+\mu)}
\left\{ q^{20j^2+j} \Mults{L}{(L-\mu)/2-4j-k,(L-\mu)/2+4j-k}{q} \right. }
\nonumber
\\ & &  \qquad \qquad  \qquad \qquad
\left. - q^{20j^2+9j+1} \Mults{L}{(L-\mu)/2-4j-k-1,(L-\mu)/2+4j-k+1}{q}
\right\} .
\end{eqnarray}
Here the variable $\mu=0,1$ is given by the parity of $L$
via $(L-\mu)/2 \in \Integer$.
Multiplying by the factor $q^{L^2/2}$ and taking the
thermodynamic limit using the result
\begin{equation}
\lim_{N\to\infty}
\Mult{2N}{N-a,N-b}{q} = \frac{1}{(q)_{\infty} (q)_{a+b}}
\qquad a+b\geq 0,
\end{equation}
yields
\begin{equation} q^{\mu/2}  \; \sum_{k=0}^{\infty}
\frac{q^{2k(k+\mu)}}{(q)_{2k+\mu}}
 \times \frac{1}{(q)_{\infty}}
\sum_{j=-\infty}^{\infty}\left\{ q^{20j^2+j} - q^{20j^2+9j+1}  \right\}
=q^{1/48+7/240}\chi_{1+\mu,1}^{(4)}(q) \: \chi_{1,1}^{(5)}(q) .
\end{equation}
This final expression indeed has the expected factorized form as
alluded to before, with the second term being the $\chi_{1,1}^{(5)}$
character corresponding to a $c=7/10$ conformal field theory.

We now turn to the fermionic LHS of (\ref{E8id}).
After replacing $q$ with $1/q$, applying (\ref{inv2})
and multiplying by the factor $q^{L^2/2}$, the fermionic sum
takes the form
\begin{equation}
\sum_{\svec{n}} q^{(\svec{n}-L\,\svec{e}_1/2)^T
C_{\mbox{\tiny E}_8}^{-1}
(\svec{n}-L\,\svec{e}_1/2) }
\prod_{i=1}^8 \Mult{n_i + m_i}{n_i}{q}
\end{equation}
where the sum is an unrestricted sum over the
components of $\vec n$ and we regard
the components $m_i$ as given in terms of
$n_i$ by the TBA system (\ref{TBA}).
We split the sum into two parts with the restrictions
\begin{equation}
n_2+n_4+n_8=\mu, 1-\mu \pmod{2}
\end{equation}
where $\mu=0,1$ gives the parity of $L$.
After making the shifts
$n_1\to L/2 -n_1-\ell$ and $n_1\to L/2 -n_1-1/2-\ell$,
respectively, where
\begin{equation}
\ell=(3n_2+4n_3+5n_4+6n_5+4n_6+2n_7+3n_8)/2
\end{equation}
the fermionic sum can be written as
\begin{eqar}
&&\sum_{\svec{n}\atop n_2+n_4+n_8=\mu \pmod{2}}
q^{(2n_1)^2/2+\overline{\svec{n}}^T
C_{\mbox{\tiny E}_7}^{-1} \,
\overline{\svec{n}} } \:
\Mult{L/2+3(2n_1)/2-\ell}{2(2n_1)}{q}\;
\prod_{i=2}^8
\Mult{n_i + \overline{m}_i}{n_i}{q}\nonumber\\
&+&\sum_{\svec{n}\atop n_2+n_4+n_8=1-\mu \pmod{2}}
q^{(2n_1+1)^2/2+\overline{\svec{n}}^T
C_{\mbox{\tiny E}_7}^{-1} \,
\overline{\svec{n}} } \:
\Mult{L/2+3(2n_1+1)/2-\ell}{2(2n_1+1)}{q}\;
\prod_{i=2}^8
\Mult{n_i + \overline{m}'_i}{n_i}{q}.
\end{eqar}
Here $\overline{m}_i$, $\overline{m}'_i$ satisfy the TBA system
\begin{equation}
\overline{\vec n} + \overline{\vec m} =\frac{1}{2}
\left ( {\cal I}_{\mbox{\scriptsize E}_7}
\overline{\vec m} + (2\overline{n}_1-1) \vec e_2 +\vec e_i \right)
\qquad i=2,\ldots,8,
\label{TBA_E7}
\end{equation}
where $\overline{n}_1=2n_1, 2n_1+1$ is even or odd, respectively.
Combining the two sums into one sum and replacing $\overline{n}_1$ with
$n_1$ now gives
\begin{equation}
\sum_{\svec{n}\atop n_1+n_2+n_4+n_8=\mu \pmod{2}}
q^{n_1^2/2+\overline{\svec{n}}^T
C_{\mbox{\tiny E}_7}^{-1} \,
\overline{\svec{n}} } \:
\Mult{L/2+3n_1/2-\ell}{2n_1}{q}\;
\prod_{i=2}^8
\Mult{n_i + \overline{m}_i}{n_i}{q}.
\end{equation}
Taking the limit $L\to\infty$ and equating this to the
bosonic RHS gives the identity
\begin{equation}
\sum_{\svec{n}\atop n_1+n_2+n_4+n_8=\mu \pmod{2}}
{q^{n_1^2/2}\over (q)_{2n_1}}\;
q^{\overline{\svec{n}}^T
(C_{\mbox{\tiny E}_7})^{-1}
\, \overline{\svec{n}} }\;
\prod_{i=2}^8
\Mult{n_i + \overline{m}_i}{n_i}{q}
=q^{1/20}\chi_{1+\mu,1}^{(4)}(q) \: \chi_{1,1}^{(5)}(q).
\end{equation}
This identity clearly has an A$_1\times\mbox{E}_7$ structure and is
very similar to the E$_7$ identity of Kedem {\em et al.}~\cite{Kedem2}.
Indeed, these results suggest that the $c=1/2$ character
should explicitly factor out of the LHS of
this identity, but we have been unable to do this.

\nsection{Some counting formulas}
In this last section we list some fermionic
expressions for the number of $L$-step paths on the dilute A$_3$
adjacency graph
${\cal G}_{\mbox{\scriptsize dA}_3}$, shown in figure~1c. These results
come about by realizing that in the critical
$q\to 1$ limit, the function $B_{r,s}(L,a,b)$ counts the number of paths
from $a$ to $b$ of length $L$ on
${\cal G}_{\mbox{\scriptsize dA}_3}$ \cite{WarPeaSeaNien}.
In other words, in this limit, the function
$B_{r,s}(L,a,b)= ({\cal I}_{\mbox{\scriptsize dA}_3})^L_{a,b}$, with
${\cal I}_{\mbox{\scriptsize dA}_3}$ the incidence matrix
corresponding to ${\cal G}_{\mbox{\scriptsize dA}_3}$.
(We remind the reader that $\lim_{q\to 1}
\left[{\textstyle{N \atop m_1,m_2,\ldots,m_n} } \right]_q =
\left({\textstyle{N \atop m_1,m_2,\ldots,m_n} } \right)$,
the RHS being an ordinary multinomial.)

Setting $q\to 1$ in some of our ``fermionic sum = bosonic sum''
identities (not all of which are listed in this paper) we find
\begin{eqnarray}
\lefteqn{
F_1(L) |_{q=1} = \left( {\cal I}^{\mbox{\scriptsize dA}_3}
\right)^L_{1,1}  \qquad \qquad
F_2(L) |_{q=1} = \left( {\cal I}^{\mbox{\scriptsize dA}_3}
\right)^L_{2,2} } \nonumber \\
\lefteqn{
F_7(L) |_{q=1} = \left( {\cal I}^{\mbox{\scriptsize dA}_3}
\right)^L_{1,2} \qquad \qquad
F_8(L) |_{q=1} = \left( {\cal I}^{\mbox{\scriptsize dA}_3}
\right)^{L+1}_{1,3}. }
\end{eqnarray}
As  ${\cal I}_{\mbox{\scriptsize dA}_3}$ has only four distinct
entries these results are complete.

\nsection{Summary and discussion}
In this letter we have shown directly that the dilute A$_3$
model in  regime 2$^+$ exhibits a hidden E$_8$ structure.
This was achieved by rewriting the known bosonic expressions
for the 1-dimensional configuration sums in
terms of fermionic sums involving the E$_8$ root system.
Our results confirm recent work of ref.~\cite{BazNienWar}
where an E$_8$ structure was found using the
Bethe Ansatz approach together with an appropriate string hypothesis.
As a byproduct of our work, we prove E$_8$~\cite{Kedem2} and
A$_1\times\mbox{E}_7$ type Rogers-Ramanujan identities.

To conclude, we point out that a similar program can be carried out for
the dilute A$_4$ and A$_6$ models~\cite{WarNienSea}.
In doing so we find that these models in regime 2 exhibit E$_7$
and E$_6$ structures, respectively.
This again confirms the earlier findings of ref.~\cite{BazNienWar2}
that these two models correspond to the exceptional
$S$-matrices of Zamolodchikov and Fateev~\cite{FatZam}.
We hope to report these results together with the complete proof
of identity (\ref{E8id}) in a future publication.

\section*{Acknowledgements}
This work is supported by the Australian Research Council.

\end{document}